\documentclass[preprint2]{aastex}

\newcommand{\sub}[1]{_{\rm #1}}

\newcommand{\oii}{{\sc [Oii]}}
\newcommand{\Mstar}{M\sub{star}}

\begin{document}

\title{The Mass Assembly and Star Formation Characteristics of Field
Galaxies of Known Morphology}
\author{Jarle Brinchmann\altaffilmark{1}}
\email{jarle@astro.ox.ac.uk}
\author{Richard S. Ellis\altaffilmark{2}}
\affil{Institute of Astronomy, Cambridge}

\altaffiltext{1}{present address: Nuclear and Astrophysical
  Laboratory, Keble Road, Oxford OX1 3RH}
\altaffiltext{2}{present address: Henry
Robinson Laboratory,MS 105-24, California Institute of Technology,
Pasadena CA 91125}

\begin{abstract}
  We discuss a new method for inferring the stellar mass of a distant
  galaxy of known redshift based on the combination of a near-infrared
  luminosity and multi-band optical photometry. The typical
  uncertainty for field galaxies with $I<22$ in the redshift range
  $0<z<1$ is a factor 2. We apply this method to a newly-constructed
  sample of 321 field galaxies with redshifts and Hubble Space
  Telescope morphologies enabling us to construct the stellar mass
  density associated with various morphologies as a function of
  redshift. We find a marked decline with time in the stellar mass
  associated with peculiar galaxies accompanied by a modest rise in
  that observed for ellipticals. The result suggests that peculiars
  decline in abundance because they transform and merge into regular
  systems. The star formation rate per unit stellar mass indicates that
  massive systems completed the bulk of their star formation before
  redshift one, whereas dwarfs continue to undergo major episodes of
  activity until the present epoch.
\end{abstract}


\section{Introduction}
\label{sec:intro}

Our understanding of the evolution of galaxies during the
important redshift range $0<z<1$ has advanced significantly in
recent years through the completion of statistically-complete
redshift surveys (Lilly et al 1995, Ellis et al 1996)
and the association of these data with
morphological information from Hubble Space Telescope (HST,
Brinchmann et al 1998, B98). This progress has been mirrored by the
formulation of detailed semi-analytic models which attempt to
reproduce the observational trends in the context of hierarchical
galaxy formation (Baugh et al 1998; Kauffmann et al 1999).

One of the most intriguing results from HST imaging and
ground-based spectroscopy is that the steep slope in the blue
number counts arises from a surprisingly abundant population of
irregular, star-forming galaxies whose local counterparts are
unclear (Glazebrook et al.\ 1995). Such galaxies contribute a negligible
amount to the local rest-frame blue luminosity density (Marzke et
al 1998) but equal the contribution of regular spirals at
redshifts $z\simeq$1 (B98). In hierarchical
models, ``morphologies'' transform according to the mass
characteristics of mergers. Recent merging leads to growth in
regular systems at the expense of lower mass irregulars (Baugh et
al 1998).

It is observationally difficult to verify this hypothesis using
traditional methods based on magnitude-limited morphological counts,
not only because galaxies transform from one kind to another but also
since they presumably brighten in optical luminosity as they merge. To
make progress we require an independent `accounting variable' capable
of tracking the likely assembly and transformation of galaxies during
the interval $0<z<1$. The colour and emission line characteristics are
transient properties and poorly-suited for this purpose (Glazebrook et
al 1999).  Ideally we seek a quantity that is readily observable
but unaffected by morphological transformations and mergers.

The dynamical or stellar mass is the obvious
choice. The former is directly connected to theoretical
predictions 
but difficult to observe except in favourably oriented
regular systems (e.g.\ Vogt et al 1997). The stellar mass is less
well-determined by theory but has the very considerable advantage
that it can be derived robustly \textit{for complete samples of
galaxies arranged morphologically} using multi-band optical and
near-infrared (near-IR) photometry
enabling a complete inventory to be made for each type at various
epochs. Since stellar mass is either conserved, or a slowly
increasing quantity, it enables us to track the fate of the
abundant $z\simeq 1$ irregular systems.

This paper is concerned with demonstrating the promise of this
technique and its initial application to a statistically-complete
sample of field galaxies of known redshift.  We have adopted a
cosmological model with $\Omega_0=0.35$, $\lambda_0=0.65$ and $h=0.65$
throughout but our results are not sensitive to this choice.


\section{Infrared Photometry for a Complete HST-based Sample}
\label{sec:infrared_photometry}

Our dataset is an extension of that introduced by B98 which resulted
from HST imaging of the CFRS and LDSS redshift surveys. The key
features are spectroscopic redshifts for a statistically-complete
sample with WFPC-2 $I_{814}$ imaging suitable for morphological
classifications, both visually and with automated classifiers (Abraham
et al 1996). By adding archival data for the Hubble Deep Field North
(HDF-N) and its flanking fields (Williams et al 1996) and the Hawaii
Deep Fields (Cowie et al 1996), we obtain a spectroscopic and
morphological sample of 498 field galaxies with $17.5<I_{AB}<22.5$.
Further details of this sample and its uniformity are discussed by
Brinchmann (1999).

Here we analyse a subset of 321 galaxies for which deep K-band
(2.2$\mu$m) photometry has been obtained from the 3.8m UK Infrared
Telescope and the literature. Imaging was performed by the authors
using IRCAM3 during three photometric nights 21--24/5 1998 using a K
filter.  Additional data came from Menanteau et al's (1999) $HK'$
survey of the Medium Deep Fields, Dickinson's (unpublished) $K$
exposures of HDF-N and Barger et al's (1999) $HK'$ imaging of the HDF
flanking fields and some original CFRS $K^\prime$ photometry (Hammer
et al.\ 1997). Total magnitudes \& aperture colours have been
re-measured in a consistent manne; the various IR filter differences
are taken into account in the SED fitting algorithm discussed below.

\section{An Improved Estimator of Stellar Mass for Distant
Galaxies}
\label{sec:method}

\begin{table}
\leavevmode
\begin{deluxetable}{ll}
  \small
    \tablecaption{Grid of Model Spectral Energy Distributions
      \label{tab:tab1}} 
    \tablewidth{8.8cm}
    \tablenum{1}
    \tablehead{\colhead{Parameter} & \colhead{Range}}
    \startdata
Age & $10^8$--$2\times 10^{10}$ years in steps of 0.1 in log
age.\\ 
SFR & $\tau=$0.1, 0.5, 1.0, 2.5, 5.0 Gyr and constant\\
Metallicity & $Z=0.02$, $Z=0.004$\\
$E(B-V)$ & 0.0, 0.2, 0.4 --- Calzetti (1997) law\\ 
    \enddata
\end{deluxetable}
\end{table}

Many authors have pointed out that the infrared luminosity of a galaxy
is a more robust estimator of its stellar mass than that derived at
optical wavelengths. The key role of $K$-band magnitudes was
emphasised as a possible test of the merging hypothesis by Broadhurst,
Ellis \& Glazebrook (1992) and this has motivated numerous
infrared-based deep surveys (Cowie et al 1996, Cohen et al 1999). The
$K$ luminosity distribution of distant galaxies was discussed in
support of the hierarchical picture by Kauffmann \& Charlot (1998) who
demonstrated its insensitivity to the previous star formation history.
The main uncertainty in the inferred stellar masses arises from the
age of the stellar population (Rix \& Rieke 1993).

The basis of our new method is to apply a correction for this
uncertainty. We adopt a grid of evolutionary synthesis models with
simple star formation histories (Bruzual and Charlot 2000, BC,
Table 1) and select the most appropriate evolutionary history by
optimal fitting of the optical+IR photometry. Extinction, age and
metallicity are allowed to vary within the bounds tabulated in a
manner similar to that adopted for photometric redshift determination
(c.f. Giallongo et al 1998) and the mass is estimated by normalising
the best fit SED to the observed $K$-band magnitude.

\begin{figure*}[tbp]
\centerline{\includegraphics[width=0.9\textwidth]{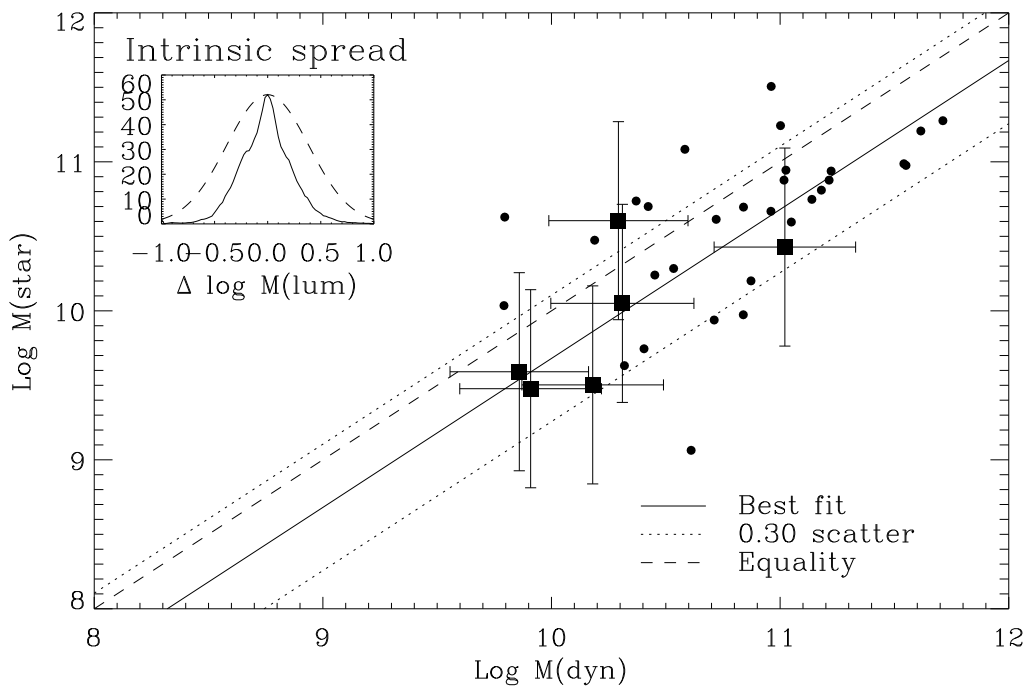}}
\figcaption[fig1.ps]{A comparison of dynamical mass
  estimates within three scale-radii ($M_{\rm dyn}=V^2 R/G$) with the
  stellar mass estimates detailed in the text. Symbols with error
  bars: six galaxies from Vogt et al.\ (1997) from the HDF flanking
  fields; these have magnitudes measured on a uniform system. Filled
  circles: 30 galaxies from Boselli et al.\ (1997). The inset
  histogram shows the average mass distribution from the Monte
  Carlo simulations described in the text and compares this with a
  Gaussian of width 0.3 in $\log \Mstar$.
\label{fig:fig1}}
\end{figure*}

Clearly there are some significant assumptions in this approach. Our
approach here is to introduce the method and its potential whilst
stressing that independent verification is clearly desirable.  We
assume (a) that the IMF does not vary systematically with
morphological type, and b) that the BC code gives reasonable
eastimates of the near-IR fluxes for a range of stellar populations.
Both assumptions are, to some extent, supported by data for external
galaxies (e.g. Kennicutt 1998) and comparisons of spectral synthesis
codes by Charlot, Worthey \& Bressan (1996).  For much of the
discussion, relative stellar masses will suffice but where absolute
masses and estimates of star formation rates are required we adopt a
Salpeter IMF with lower and upper mass cut-offs of 0.1 $M_{\odot}$ and
125 $M_{\odot}$.

To assess the \emph{random} errors arising from our method, we
propagate magnitude errors through the fitting procedure using a Monte
Carlo method. This gives an \emph{intrinsic} error on the mass
estimates with a median error of $0.20$dex in $\log \Mstar$ (see inset
histogram in Figure~\ref{fig:fig1}). Further errors might arise due to
our choice of exponentially declining star formation histories, and to
quantify this we have explored the sensitivity of the method to a
secondary burst of SF and find that this can introduce scatter of
$\sim 0.2$ in $\log \Mstar$ with no sign of systematic offsets.  Taken
together this gives an estimate of the theoretical uncertainty in the
mass estimates of $\approx 0.3$dex in $\log \Mstar$.


One potentially useful test of our method is a comparison of our
stellar mass estimates with those determined dynamically using
rotation curves. We currently have a small overlap of 6 galaxies with
Vogt et al (1997) and assuming that these have accurately estimated
terminal velocity speeds we can estimate the dynamical mass within
three scale-radii using $M_{\rm dyn}=V^2 R/G$.  The comparison of this
mass estimate with the stellar mass is shown in Figure~\ref{fig:fig1}.
The same exercise has been performed with the local sample of Boselli
et al (1997) although we expect a larger scatter here as the
photometric scales have not been accurately aligned.
Figure~\ref{fig:fig1} shows both masses correlate with a scatter
consistent with earlier estimes. Of course, the relationship between
dynamical and stellar mass depends on the physical properties and is
of intrinsic interest.  The mean offset corresponds to $M_{star}$ =
0.3-0.5 $M_{dyn}$. Clearly it will be instructive to construct Figure
1 for a larger sample.

\section{Mass Assembly and Specific Star Formation Rates}
\label{sec:massflow}

Although considerable progress has been made in delineating the
star formation history of field galaxies (e.g.\ Madau, Pozzetti \&
Dickinson 1998), such probes remain uncertain particularly
as there is not yet good agreement on what represents a
reliable diagnostic. Moreover, it is clear from the considerable
diversity of predictions in semi-analytic models (c.f.\ Baugh et al
1998) that cosmic star formation histories cannot
precisely test hierarchical theories of galaxy formation. Whilst
considerable effort has been invested in incorporating realistic
aspects of star formation and feedback into numerical models of
galaxy evolution, here we explore an alternative route, viz. using
our method to bring the empirical data closer to the fundamental
aspects of hierarchical theory by constructing the redshift
dependence of the stellar mass function.

To complement the co-moving volume-averaged star formation history
$\rho_{\ast}(z)$, we seek to construct its stellar mass equivalent
$\rho_{m}(z)$ and particularly its dependence on galaxy
morphology. With sufficient data we could also construct the
stellar mass function $\Phi(m)$, analogous to the luminosity
function, using a $1/V_{max}$ formalism. Its consideration allows
us to discuss the important question of possible incompleteness
which would otherwise underestimate the stellar mass density
$\rho_m(z)$.

Incompleteness can arise in various ways. The most critical is that
galaxies beyond our $I_{814}$ limit in a given redshift range
contribute to the infrared flux and hence the stellar mass density.
Ultimately we seek a $K$-limited version of our current sample from
which galaxies can, to first order, be selected to a fixed stellar
mass limit at a given redshift. Meanwhile, to assess this
incompleteness, we utilise the form of the K-band galaxy luminosity
function (LF) obtained by combining the recent infrared survey of
Cohen et al (1999) with the $K$-band LF of Loveday (1999).

\begin{figure*}[tbp]   
\centerline{\includegraphics[width=0.8\textwidth]{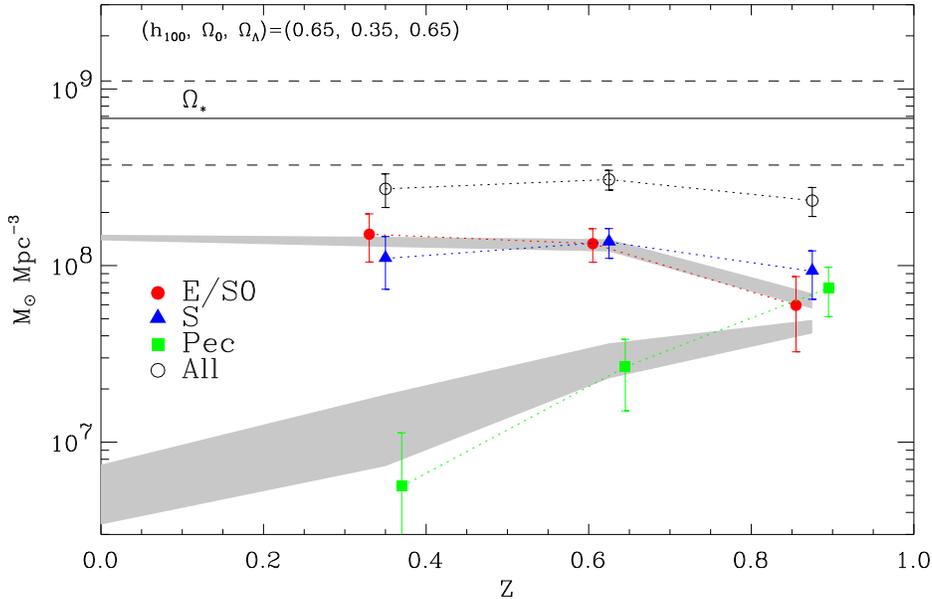}}
\figcaption[fig2.ps]{The integrated stellar mass density for
  galaxies with $10.5<\log \Mstar < 11.6$ as a function of redshift
  and visual morphology. The shaded regions show the predictions of
  the mass density in peculiar and elliptical galaxies from the
  simple merging models described in the text.\label{fig:fig2}}
\end{figure*}

Assuming a fixed mass/light ratio of $\Mstar/L_K$=0.8, this LF
suggests that if we restrict ourselves to galaxies with stellar masses
in the range $10.5<\log \Mstar<11.6$, we will be $>80\%$ complete in
$\rho_m$ across the interval $0.2<z<1.0$, with the greatest
incompleteness in the highest redshift bin. In the subsequent
discussion we have restricted our analysis to this mass range and
taken the incompleteness corrections and their uncertainty into
account.

\begin{figure*}[tbp]
\centerline{\includegraphics[width=0.75\textwidth]{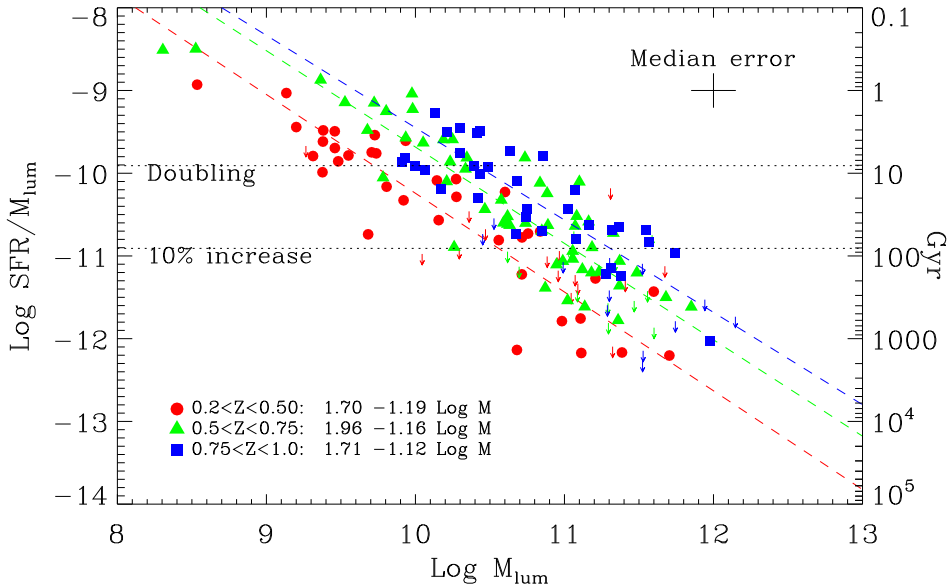}}
\figcaption[fig3.ps]{The specific star formation rate, $R$, for
  the galaxies in the sample. The three lines show orthogonal
  least-squares fit to the three redshift ranges indicated. The
  fit parameters are shown in the lower left corner. The right
  hand panel shows the doubling time in Gyr, assuming constant
  star formation. Arrows show 2$\sigma$ upper-limits for galaxies
  with no detected \oii\ emission.\label{fig:fig3}}
\end{figure*}

Figure~\ref{fig:fig2} shows the comoving stellar mass density in
galaxies partitioned by morphology as a function of redshift with
error bars derived by bootstrap resampling methods. The solid
horizontal line shows the total local stellar mass density inferred
from a recent review by Fukugita, Hogan \& Peebles (1998) adjusted to
our IMF. The most significant feature is the remarkable decline in
mass density associated with galaxies of irregular morphology. This is
complemented by a modest increase in that associated with regular
galaxies, particularly spheroidals, ensuring that the mean stellar
mass density remains approximately constant at all redshifts probed.
Inspection of the mass function for high redshift irregulars
(Brinchmann 1999) shows a significant number harbour stellar masses as
large as those for the most massive regular galaxies suggesting that
these simply transform into the latter systems perhaps as a result of
the onset of more regular star formation patterns. However, the
decline in the irregular galaxy luminosity density may also be driven
by merging, causing a gradual increase in the mass density
associated with regular galaxies.

What rate of merging is necessary to explain Figure~\ref{fig:fig2}? By
accounting for evolution in terms of stellar mass, rather than blue
luminosity, we can consider this issue using recent estimates of the
redshift-dependent merger fraction (Le F{\`e}vre et al 2000, and
references therein). Adopting a merger fraction $\propto (1+z)^{3.4}$,
Brinchmann (1999) finds a good fit to the morphologically-dependent
redshift distribution and luminosity functions if 5-10\% of the
present population of E/S0s formed via mergers over $0<z<1$. This
solution and its uncertainty is illustrated by the shaded area in
Figure~\ref{fig:fig2} where the upper locus refers to a model where
the merger progenitors are assumed to be of average mass and with
colours similar to local Sbc galaxies over the entire redshift range.
However the data indicate that peculiar galaxies become less massive
at low redshift, hence we show a model where merger progenitors at
$z=0.5$ are 30\% less massive than those at $z=1$ as the lower locus.
The total mass density is normalised to the observed amounts in each
redshift range, and the model chosen such that it reproduces the local
morphologically-split LFs from Marzke et al.\ (1998). The increase
with time in the E/S0 mass density is well matched in both models
whereas the decrease in the peculiar mass fraction is overestimated.
Converting some peculiar galaxies into spirals (whose mass density was
assumed to be constant in the foregoing) would improve the fit.

A further important result can be extracted from
Figure~\ref{fig:fig2}, namely that the total stellar mass density in
massive galaxies is approximately constant over the interval
$0.2<z<1$. This, in turn, implies that most of the stellar content of
these galaxies must have formed at higher redshift. At first sight
this seems a surprising result given, over the same redshift range,
the volume-averaged star formation rate remains high (Madau et~al.\ 
1998). Our results can be more easily understood by
considering the contribution that star formation makes to the total
mass of a galaxy.

Consider the {\it specific star formation rate} $R= SFR/\Mstar$ which
defines a useful measure of the rate at which new stars add to the
assembled mass of a galaxy. For massive galaxies, a much larger $R$ is
needed to make a fractionally significant mass contribution than would
be the case for a low mass system. Figure~\ref{fig:fig3} shows $R$ as
a function of stellar mass for the CFRS galaxies in our survey (Hammer
et al.\ 1997). Different symbols relate to three redshift ranges,
arrows show 2$\sigma$ upper limits. Star formation rates were derived
using the luminosity observed in the [O II]3727 \AA\ emission line
using (Guzman et al.\ 1997)
\begin{equation}
  \label{eq:sfr_conv}
  SFR(M_{\odot} {\rm yr}^{-1}) = 
  10^{-12.4-0.4(M_B-M_{B,\odot})} EW({\mbox{\oii}}),
\end{equation}
where $M_B$ is the blue luminosity. Note that Figure~\ref{fig:fig3}
makes a strong statement about the upper envelope of the distribution
where incompleteness has no effect.

The derivation of stellar masses for each galaxy in our survey enables
us to ask when galaxies of a particular mass range formed most of
their stars. The right hand axis in Figure~\ref{fig:fig3} shows the
doubling time for the stellar mass of a galaxy assuming constant star
formation, with the horisontal lines indicating the value of $R$
required to double and increase by 10\% the stellar mass of a galaxy
from $z=1$ to the present.  Clearly massive galaxies for the bulk of
their stellar content at $z \gtrsim 1$.
At lower redshift, the pattern continues for lower mass galaxies
consistent with the "downsizing" picture first introduced by Cowie et
al (1996) and our findings are also in very good agreement with Guzman
et al.\ (1997) when the differences in mass estimation are taken into
account.  Given our incompleteness, dwarf galaxies may well form over
a wide redshift range since we cannot probe such mass ranges at all
redshifts.

A potential problem with this argument would be massive amounts of
obscured star formation, since we have made no dust correction beyond
the local calibration.  We cannot rule this out conclusively, but
observe that only few galaxies in our sample show significant far-IR
emission (Flores et al.\ 1999), and even a constant star formation
rate of $10M_{\odot}/\mbox{yr}$ will only cause a 10\% increase in the
mass density over the redshift and mass range we here consider.

\section{Conclusions}
\label{sec:summary}

We have presented a new method for determining reliable stellar masses
for field galaxies of known redshift and applied it to a uniform
sample of 321 galaxies with HST morphologies. We demonstrate via
comparisons and simulations that our method yields stellar masses with
a precision of $\log \Delta \Mstar \approx 0.3$dex, although further
calibrations and comparisons with dynamical estimates would be
valuable. The main advantage of our method is that, with few
assumptions, it provides stellar masses with greatly reduced
observational requirements and with no bias against galaxy type or
orientation. It is therefore particularly well-suited for considering
the mass density of galaxies of different types as a function of
redshift.

Whereas only a small proportion ($<10$\%) of massive field
spheroidal galaxies can have formed by merging between $z=0$ and
$z=1$, this most likely arises from a significant demise in the
mass density of high-$z$ peculiar galaxies. The transformation of
irregulars into more massive regular systems is shown to be in
reasonable agreement with the evolution of the merger fraction
found independently from pair statistics.

We find no evidence for any change in the mass density of spiral
galaxies, suggesting the merging of spiral galaxies into
elliptical galaxies between $z=1$ and the present cannot be a
dominant process. Moreover, the total stellar mass density hardly
increases over the range sampled. Both results support the
contention that most massive galaxies formed their stars prior to
or around $z=1$. Lower mass galaxies remains active to much more
recent epochs consistent with the "downsizing" picture introduced
by Cowie and collaborators.

\acknowledgements

JB acknowledges receipt of a PPARC fellowship and an Isaac Newton
Fellowship. We would like to thank Bob Abraham, Shaun Cole, George
Efstathiou, John Huchra, Masataka Fukugita, Guinevere Kauffmann,
Piero Madau, Gabriela Mall{\'e}n-Ornelas and Simon White for
suggestions and constructive criticisms and Judy Cohen for supplying
the Caltech Redshift Survey catalogue.

\end{document}